\documentclass[9pt,twocolumn,twoside]{opticajnl}

\journal{ao} 

\setboolean{shortarticle}{false}

\usepackage{subfigure}
\usepackage{lineno}

\title{Robustness of optic-fiber-based weak value amplification against amplitude-type noise}

\author[1]{Han Wang}
\author[1,*]{Jingzheng Huang}
\author[1]{Chaozheng Huang}
\author[1]{Hongjing Li}
\author[1]{Guihua Zeng}

\affil[1]{School of Electronic Information and Electrical Engineering, State Key Laboratory of Advanced Optical Communication Systems and Networks, Institute of Quantum Sensing and Information Processing, Shanghai Jiao Tong University, Shanghai 200240, China}

\affil[*]{Corresponding author: jzhuang1983@sjtu.edu.cn}

\begin{abstract}
Experiments based on free space platform have demonstrated that the weak value amplification (WVA) technique can provide high sensitivity and precision for optical sensing and metrology. To promote this technique for real-world applications, it is more suitable to implement WVA based on optical fiber platform due to the lower cost, smaller scale and higher stability. In contrast to the free space platform, the birefringence in optical fiber is strong enough to cause polarization cross talk, and the amplitude-type noise must be taken into account. By theoretical analysis and experimental demonstration, we show that the optic-fiber-based WVA is robust in presence of amplitude-type noise.
In our experiment, even the angular misalignment on optical axes at the interface reaches $0.08 rad$, the sensitivity loss can be maintained less than $3dB$. 
Moreover, the main results are valid to a simplified detection scheme that recently proposed, which is more compatible with the future design of optical-fiber based WVA. 
Our results indicate the feasibility of implementing WVA based on optical fiber, which provide a possible way for designing optical sensors with higher sensitivity and stability in the future.
\end{abstract}

\setboolean{displaycopyright}{true}

\begin{document}

\maketitle

\section{Introduction}
Three decades after the concept had been proposed, weak value amplification (WVA)\cite{Aharonov1988} has recently become a useful technique for optical metrology\cite{Hosten2008,Dixon2009,Brunner2010,Xu2013,Li2018,Dziewior2019}. Taking the advantage of suppressing certain kinds of technical noises and practical imperfections\cite{Jordan2014}, WVA can efficiently improve the signal-to-noise ratio in practice, therefore provide higher sensitivity comparing to the standard techniques\cite{Viza2015,Alves2017,Xu2020}. However, most of the proof-of-principle WVA experiments were designed and performed on free-space platforms, e.g Refs.\cite{Hosten2008,Dixon2009,Xu2013,Dziewior2019}. It is hard to directly transplant them to real-world applications due to their cost, scale and stability. 

A possible way to solve this problem is to replace the free-space devices in WVA experiments by their optical fiber counterparts, thanks to the rapidly developed fiber optic sensor technology\cite{FOS2002}. To achieve this goal, it is necessary to analyze the negative effects caused by the birefringence, which is negligible in free space but inevitable in optical fiber\cite{Tsubokawa1988,VanWiggeren1999,Xu2009,fang2013coupled}. Because the birefringence in optical fiber is strong enough to cause polarization cross talk, one of the most important effect should be taken into account is the amplitude-type noise effect\cite{Dandridge1981,Handbook2009}, where the noise intensity is proportional to the amplitude of the signal, instead of the intensity-type noise that being considered in the free-space WVA\cite{Huang2018}. Recently, a low noise phase measurement of a fiber optic sensor conducted via weak value amplification is experimentally demonstrated to have a flat and wideband frequency response from 0.1 Hz to 10 kHz\cite{Liu:22}. In addition, a WVA-enhanced hydrophone using a 0.8 meter long optical fiber was experimentally demonstrated\cite{Luo2020}, which was further improved to obtain a high-performance fiber optic interferometric hydrophone with high sensitivity and wide-frequency response\cite{9481320}. However in the absence of analyzing the amplitude-type noise effect, the feasibility for a more general case, such as applying longer optical fiber or other WVA schemes\cite{Huang2019}, is not yet clear.

In this work, we theoretically and experimentally analyze the amplitude-type noise effect in WVA.
In contrast to the scheme applied in Ref.\cite{Luo2020}, we concern on the WVA schemes with assistance of spectral analysis, and all of the four approaches\cite{Huang2019} are exemplified. After building up the theoretical model and performing numerical simulations, we experimentally verified the effect of amplitude-type noise on the sensitivity of BWVA, and the feasibility of using the spectral split detection to replace optical spectrum analyzer.

This paper is organized as follows: in Sec.2, after a brief review on the preliminaries of WVA, the effect of amplitude-type noise in optical-fiber-based WVA was analyzed in theory by a generalized theoretical model along with numerical simulations. The experimental demonstrations followed by discussions are made in Sec.3, and the final conclusion remarks are made in Sec.4.

\section{Theory}
\subsection{Preliminary}
\begin{figure}[ht!]
\centering\includegraphics[width=5cm]{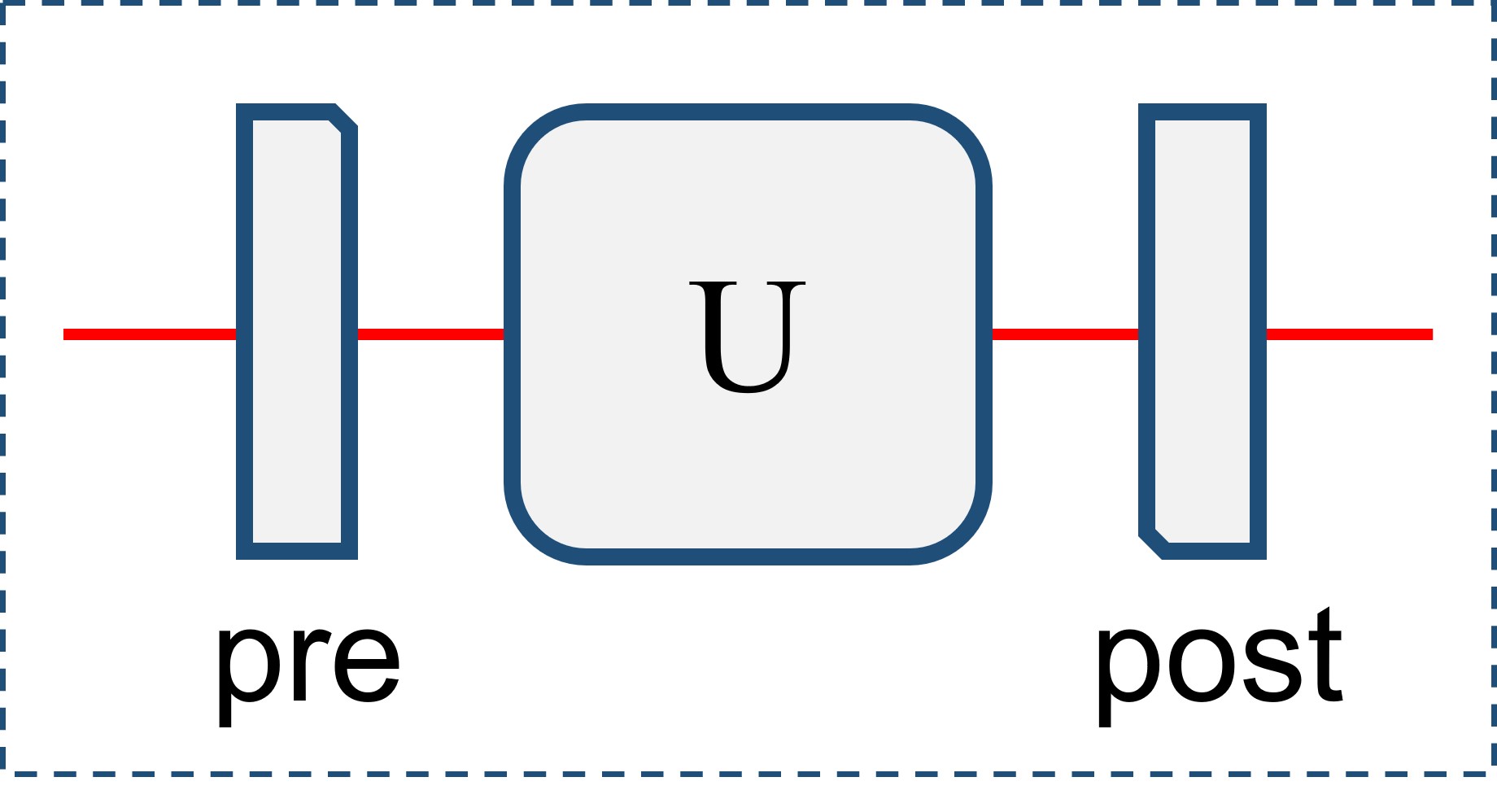}
\caption{The weak measurement process without noise.}
\label{wmmodel}
\end{figure}

We are interested of the typical weak measurement model that involves a two-level system and a pointer with continuous variable of p\cite{Jozsa2007}. As is shown in Fig.\ref{wmmodel}, the system and pointer are firstly prepared in states $|{\phi_{in}}\rangle$ and $|\psi_{in}\rangle=\int dpf(p)p$  respectively, and the total initial state is $|\varphi_{in}\rangle=|\phi_{in}\rangle\otimes|\psi_{in}\rangle$. The followed interaction between the system and pointer is described by $U=e^{-ig\hat{A}\otimes\hat{p}}$, where $\hat{A}$ is a Pauli operator acting on the system, $\hat{p}$ is the momentum operator acting on the pointer, and $g$ is a very small coupling strength which is the parameter of interest. Afterward, the system is postselected by $|{\phi_{f1}}\rangle$ and $|{\phi_{f2}}\rangle$, and the probability distribution of the pointer becomes:
\begin{equation}
\resizebox{.5\hsize}{!}{$
\begin{aligned}
    P_{f1}(p)&=\left | \left \langle \phi _{f1}|e^{-ig\hat{A}\otimes\hat{p}}|\varphi _{in} \right \rangle \right |^{2}, \\
    P_{f2}(p)&=\left | \left \langle \phi _{f2}|e^{-ig\hat{A}\otimes\hat{p}}|\varphi _{in} \right \rangle \right |^{2}
\end{aligned}$}
\end{equation}

\begin{figure}[ht!]
\centering
    \subfigure[]{\includegraphics[width=0.3\textwidth]{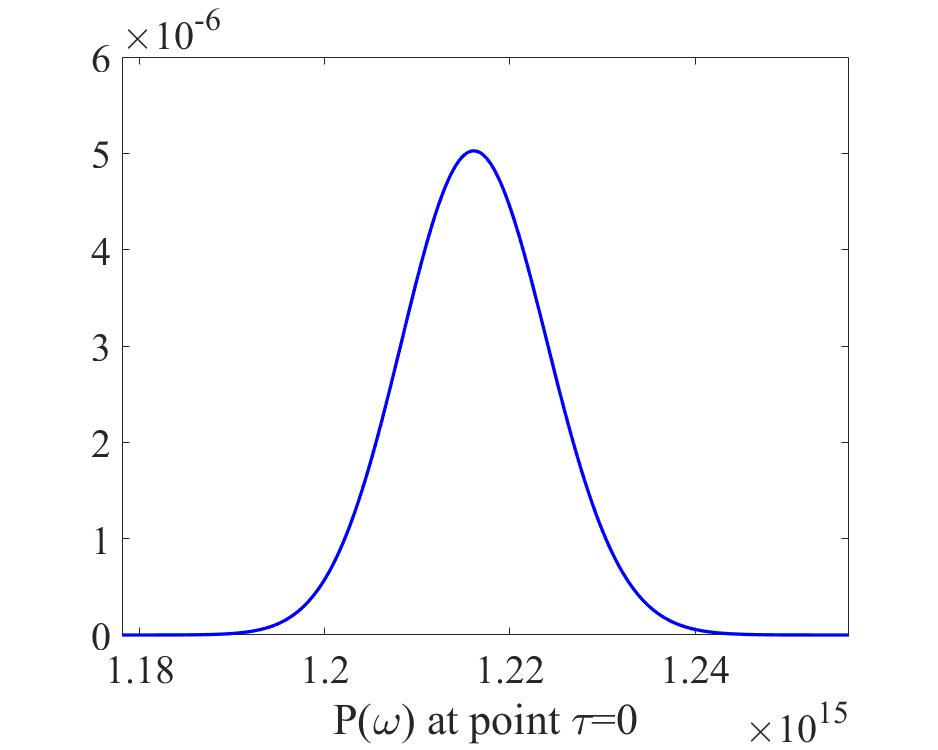}} 
    \subfigure[]{\includegraphics[width=0.3\textwidth]{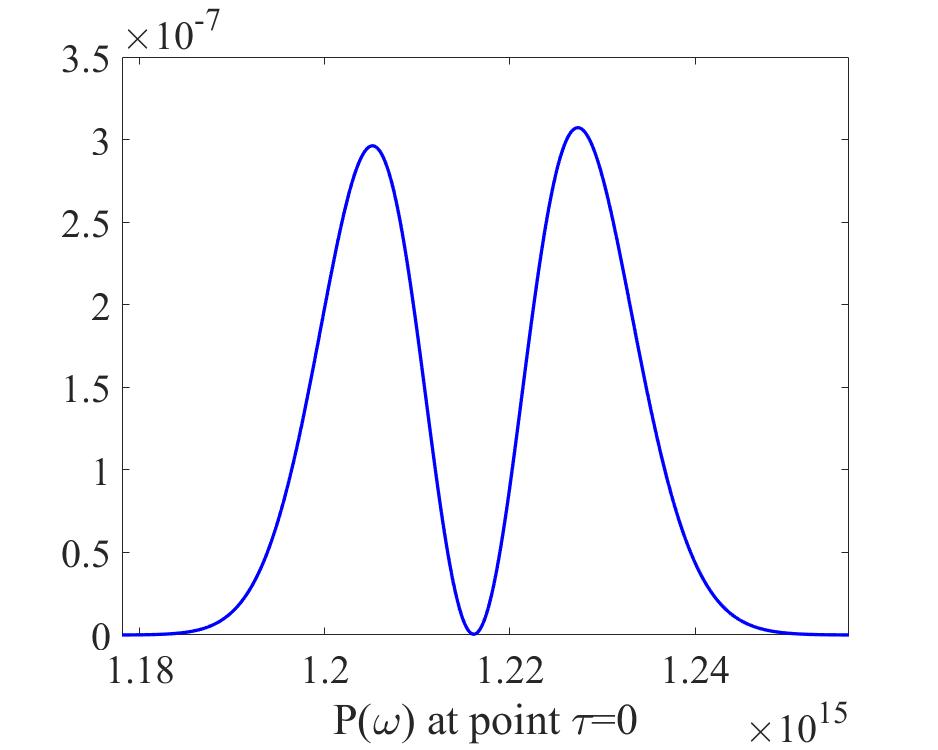}} 
    \subfigure[]{\includegraphics[width=0.3\textwidth]{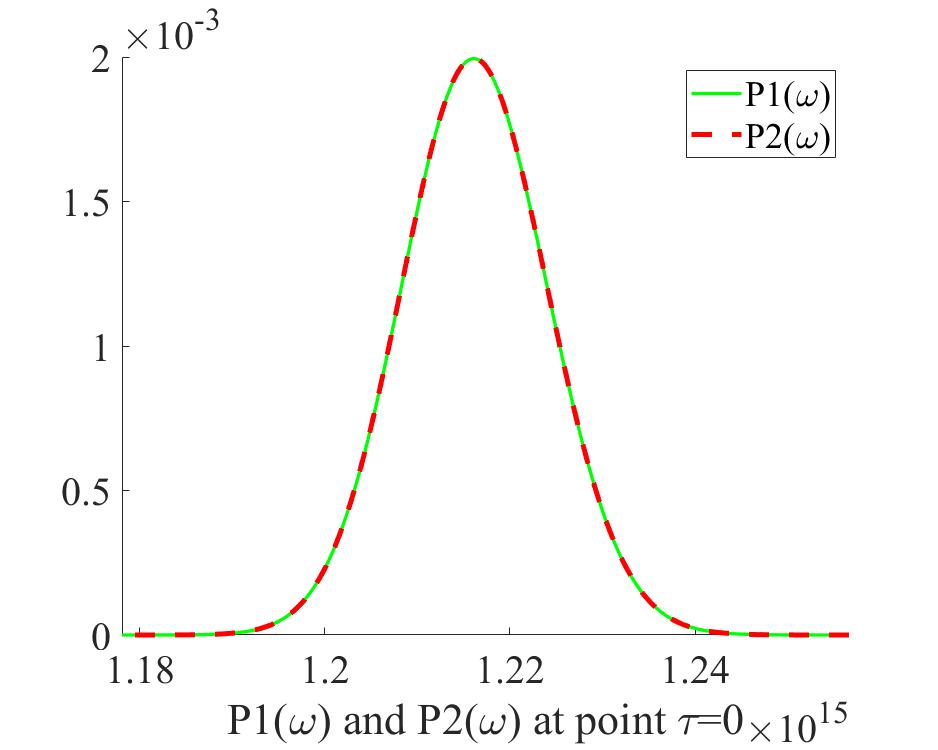}} 
    \subfigure[]{\includegraphics[width=0.3\textwidth]{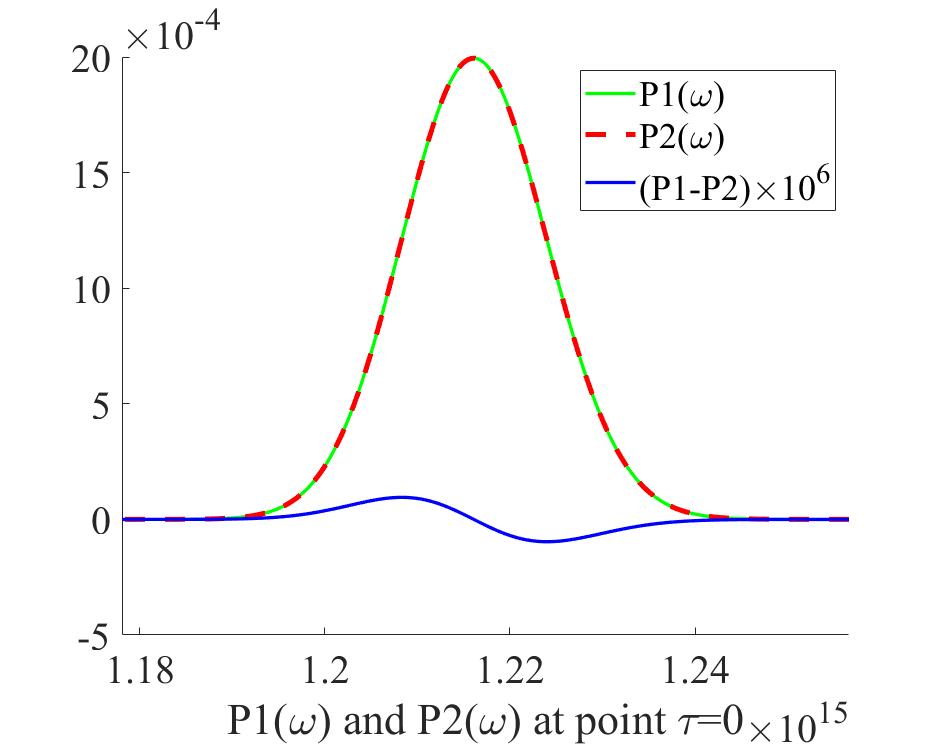}} 
\caption{Numerical simulations on the spectrum of WVA approaches without noise: (a)SWVA, (b)BWVA, (c)JWVA, (d)DWVA.}
\label{WVANoNoise}
\end{figure}

In the application of optical sensing, we adopt the optical polarization as the system, as well as the optical spectrum as the pointer, the unitary becomes:$U=e^{-i\tau\hat{A}\omega}$, where $\omega$ is the optical frequency, $\hat{A}$ is the Z-Pauli matrix and $\tau$ is a ultrasmall time delay\cite{Brunner2010}. By choosing  $|{\phi_{in}}\rangle=(|H\rangle+|V\rangle)/\sqrt{2}$, where H and V denotes the horizontal and vertical polarization states respectively. The value of $\tau$ can be revealed by the shift of the average value of $\omega$, which depends on the applied WVA approach, namely standard WVA (SWVA)\cite{Brunner2010}, biased WVA (BWVA)\cite{Zhang2016}, joint WVA (JWVA)\cite{Strubi2013} or dual WVA (DWVA)\cite{Huang2019}:
\begin{equation}\label{eq:delta_omega}
	\begin{array}{lll}
	SWVA:\begin{cases}
		\delta_{\omega}=\frac{\omega_{f1}}{\overline{P}_{f1}}-\omega_0,\\
		|\phi_{f1}\rangle= e^{i\epsilon}|H\rangle - e^{-i\epsilon}|V\rangle,
		\end{cases}\\
	BWVA:\begin{cases}
		\delta_{\omega}=\frac{\omega_{f1}}{\overline{P}_{f1}}-\omega_0, \\
		|\phi_{f1}\rangle= e^{i(1-\frac{\omega}{\omega_0})\epsilon}|H\rangle - e^{-i(1-\frac{\omega}{\omega_0})\epsilon}|V\rangle,
		\end{cases}\\
	JWVA:\begin{cases}
		\delta_{\omega}=\frac{\overline{P}_{f2}\omega_{f1}-\overline{P}_{f1}\omega_{f2}}{\overline{P}^2_{f1}-\overline{P}^2_{f2}},\\ 
		|\phi_{f1,f2}\rangle= e^{i\epsilon}|H\rangle \pm i e^{-i\epsilon}|V\rangle,
		\end{cases}\\
	DWVA:\begin{cases}
		\delta_{\omega}=\frac{\int \omega[P_{f1}(\omega)-P_{f2}(\omega)]^2d\omega}{\int [P_{f1}(\omega)-P_{f2}(\omega)]^2d\omega},\\
		|\phi_{f1,f2}\rangle= e^{i(1-\frac{\omega}{\omega_0})\epsilon}|H\rangle\pm i e^{-i(1-\frac{\omega}{\omega_0})\epsilon}|V\rangle,
	    \end{cases}
    \end{array}
\end{equation}
where $\omega_{fn}\equiv\int \omega P_{fn}(\omega)d\omega$, $\overline{P}_{fn}\equiv\int P_{fn}(\omega)d\omega$ with $n=1,2$, $\omega_0$ is the average optical frequency of the initial state, and $\epsilon$ is a small wavelength-independent phase. The output spectra of different WVA approaches without noise are given in Fig.\ref{WVANoNoise}. 

A main advantage of the WVA technique for parameter estimation is that a high sensitivity can be obtained. By defining the sensitivity of WVA by $S_{\omega}\equiv\frac{\Delta\delta_{\omega}}{\Delta\tau}$, we can compare the sensitivity of the four approaches\cite{Huang2019}:
\begin{equation}\label{eq:sensitivity_ideal}
S_{\omega}\simeq\begin{cases}
2\sigma_{\omega}^2/\epsilon & ,for\ SWVA\\
2\omega_0^2/\epsilon & ,for \ BWVA\\ 
\sigma_{\omega}^2/\epsilon & ,for\ JWVA\\
2\omega_0^2/\epsilon & ,for\ DWVA
\end{cases},
\end{equation}
where $\sigma_{\omega}$ is the optical frequency variance of the initial state.

\subsection{ Principle and theoretical model of amplitude-type noise in WVA}
To date, all of the four WVA approaches have been experimentally realised. Although ultra-high sensitivity were successfully demonstrated\cite{Xu2013,Fang2016,Martinez-Rincon2016,Li2019,Huang2019}, current experiments based on free-space platform were often with large scale and relatively poor stability, which make it difficult to satisfy the requirement for real-world applications. To overcome this obstacle, integrating the WVA setup with optical fiber seems to be a potential solution. However, due to the geometrical effect of the core or the stress effect around the core, modal birefringence always exists in optical fiber, and inevitably affects the final performance of WVA. 
In particular, one of the most important effect is the amplitude-type noise\cite{fang2013coupled,Xu2009}, which is negligible in the free-space schemes. 

\begin{figure}[ht!]
\centering\includegraphics[width=8cm]{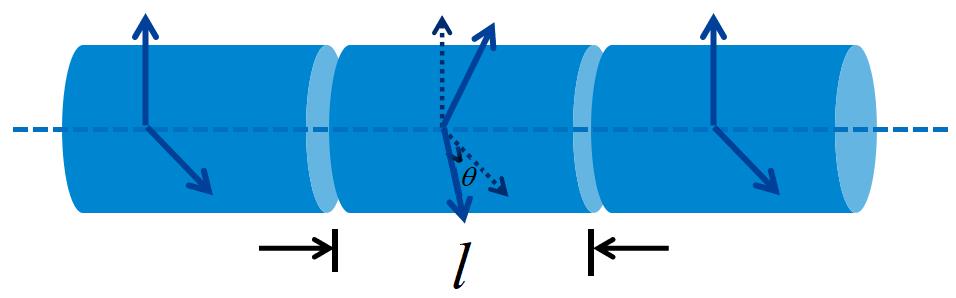}
\caption{The three-stage cascade model for a segment of optical fiber with length of $l$.}\label{fig:cascade_model}
\end{figure}

To analyze this effect, we first consider a small segment of optical fiber with length of $l$ and put it in the middle of three sections, as is shown in Fig.\ref{fig:cascade_model}\cite{Xu2009}. When this segment is under external perturbation, a birefringence axis appears at the angle of $\theta$ with respect to the preset coordinate, and the normal refractive index $n_0$ slightly changes to be $n_o$ and $n_e$. For simplicity, the intensity of cross-talk is assmued to be small enough, so that the main light intensity of each small block can be regarded as constant. Moreover, the impact of polarization dependent loss (PDL) and polarization mode dispersion (PMD) are not taken into account. We first consider an input light with linear polarization and wavelength of $\lambda$ (relating to the angular frequency by $\lambda=2\pi c/\omega$), and write it in term of Jones vector, namely $|\varphi_{in}\rangle=\begin{bmatrix}
1\\ 
0
\end{bmatrix}$. After passing through one small segment of optical fiber, the polarization state becomes\cite{Xu2009}:
\begin{equation}\label{eq:Jones}
\resizebox{1.0\hsize}{!}
{$
\begin{aligned}
\vert\varphi'_{in}\rangle&=\begin{bmatrix}\cos\theta&-\sin\theta\\\sin\theta&\cos\theta\end{bmatrix}\begin{bmatrix}e^{ikn_ol}&0\\0&e^{ikn_el}\end{bmatrix}\begin{bmatrix}\cos\theta&\sin\theta\\-\sin\theta&\cos\theta\end{bmatrix}\begin{bmatrix}1\\0\end{bmatrix}\\&=\begin{bmatrix}e^{ikn_ol}\cos^2\theta+e^{ikn_el}\sin^2\theta&\frac12e^{ikn_ol}(\sin2\theta)(1-e^{-i\frac{2\pi l}{L_b}})\\\frac12e^{ikn_ol}(\sin2\theta)(1-e^{-i\frac{2\pi l}{L_b}})&e^{ikn_ol}\cos^2\theta+e^{ikn_ol}\sin^2\theta\end{bmatrix}\begin{bmatrix}1\\0\end{bmatrix},
\end{aligned}$}
\end{equation}
where $k=2\pi/\lambda$ is the wave number, $L_{b}=\lambda/|n_e-n_o|$ is the beat length. 

We consider a optical fiber with length of L and divide it into N segments. Then after passing through the whole optical fiber, we get:
\begin{equation}\label{eq:amplitude_noise}
|\varphi _{out}\rangle=(e^{ikn_0 L}\prod_{m=1}^{N}B_{m})|\varphi_{in}\rangle+(A_{eff}e^{i\theta _{eff}})|\varphi_{in}^{\perp}\rangle,
\end{equation}
where
\begin{equation}\label{eq:Bm}
B_{m}=\sqrt{1-\sin^{2}(2\theta_m)\sin^{2}(\frac{\pi L}{L_{bm}})}
\end{equation}
and
\begin{equation}
\label{eq:Aeff}
\begin{split}
    A_{eff}e^{i\theta _{A_{eff}}} =&e^{ikn_0 l}\sum_{m=1}^{N}(\sin(2\theta _{m})\sin(\frac{\pi l}{L_{bm}})\\
&\cdot e^{i\left [ \frac{\pi}{2}-\frac{\pi l}{L_{bm}}+kn_0(L-ml) \right ]})
\end{split}
\end{equation} 
with $\theta_{m}$ and $L_{bm}$ being the birefringence axis angle and beat length of the m-th segment of the fiber, $\varphi_{in}^{\perp}$ denotes the state with polarization being orthogonal to $\varphi_{in}$, which consists of the amplitude-type noise. 

\begin{figure}[ht!]
\centering\includegraphics[width=8.5cm]{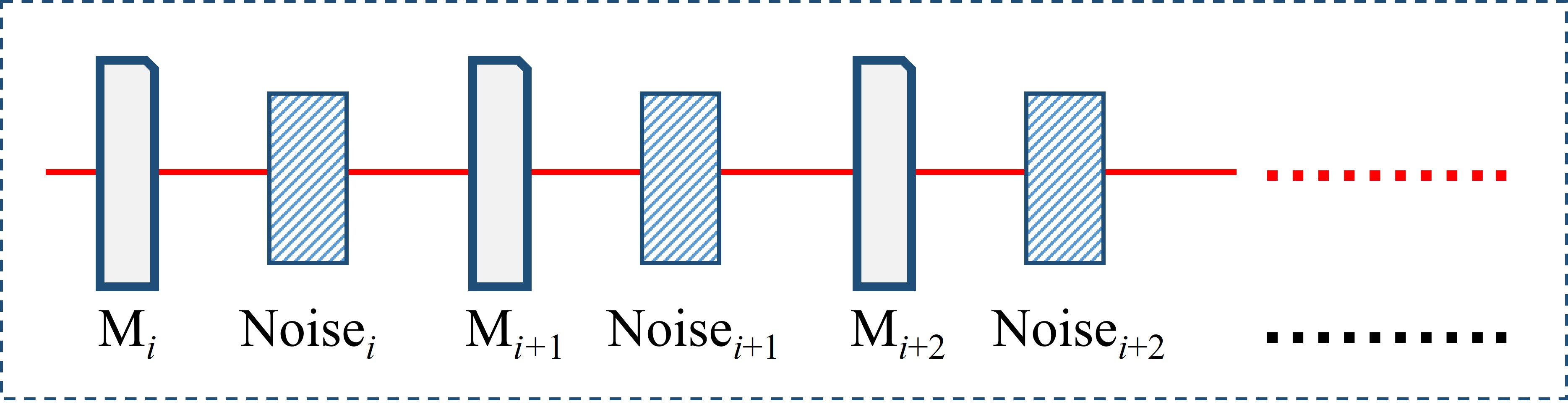}
\caption{The weak measurement process with noise in optical fiber.}
\label{Rigorousmodel}
\end{figure}

We can now move forward to explore the effect of amplitude-type noise on WVA by applying this model.  In principle, the effect corresponding to the fiber birefringence happens during the interaction process involving the parameter of interest, as is shown in Fig.\ref{Rigorousmodel}. The state evolution is given by: 
\begin{equation}
|\varphi_{in}\rangle \rightarrow e^{i\hat{A}\omega\tau_N}M_N...e^{i\hat{A}\omega\tau_2}M_2e^{i\hat{A}\omega\tau_1}M_1|\varphi_{in}\rangle,
\end{equation}
where $M_{i}(i=1,2...,N)$ represent the Jones matrix corresponding to Eq.(\ref{eq:Jones}), and $\sum\tau_i = \tau$. 

\begin{figure}[ht!]
\centering\includegraphics[width=7cm]{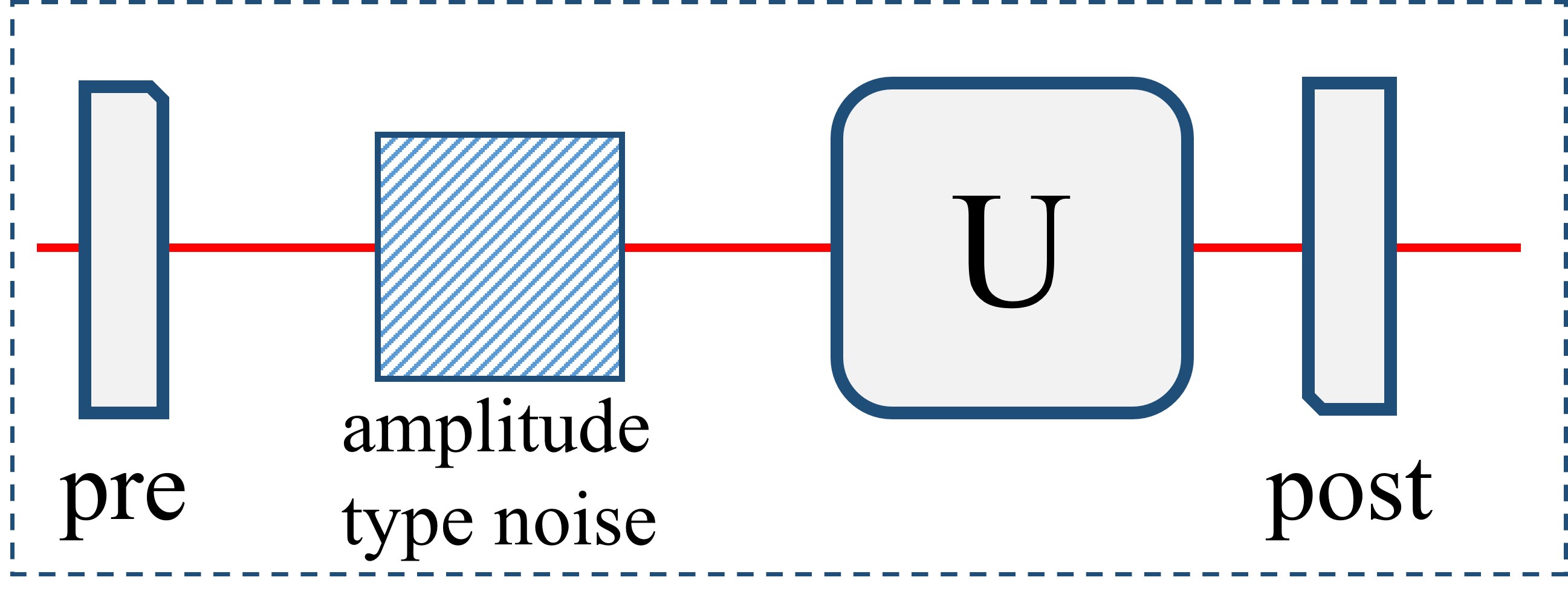}
\caption{Equivalent model for weak measurement process with noise.}
\label{fig:amplitude_noise}
\end{figure}

By expressing the first two steps we obtain:
\begin{equation}
\begin{split}
&...e^{-i\widehat A\omega\tau_3}M_3e^{-i\widehat A\omega\tau_2}M_2e^{-i\widehat A\omega\tau_1}M_1\vert\varphi_{in}\rangle\\
=&...e^{-i\widehat A\omega\tau_3}M_3e^{-i\widehat A\omega\tau_2}M_2e^{-i\widehat A\omega\tau_1}(a_1\vert\varphi_{in}\rangle+b_1e^{i\omega t_{d1}}\vert\varphi_{in}^\perp\rangle)\\
=&...e^{-i\widehat A\omega\tau_3}M_3e^{i\widehat A\omega\tau_2}M_2(a_1\vert\varphi_1\rangle+b_1e^{i\omega t_{d1}}\vert\varphi_1^\perp\rangle)\\
=&...e^{-i\widehat A\omega\tau_3}M_3e^{i\widehat A\omega\tau_2}\lbrack a_1(a_2\vert\varphi_1\rangle+b_2e^{i\omega t_{d2}}\vert\varphi_1^\perp\rangle)+\\
&\;\;\;\;\;\;\;b_1e^{i\omega t_{d1}}(a_2\vert\varphi_1^\perp\rangle+b_2e^{i\omega t_{d2}}\vert\varphi_1\rangle)\rbrack\\
=&...e^{-i\widehat A\omega\tau_3}M_3\lbrack(a_1a_2+b_1b_2e^{i\omega(t_{d1}+t_{d2})})\vert\varphi_2\rangle+\\
&\;\;\;\;\;\;\;(a_1b_2e^{i\omega t_{d2}}+b_1a_2e^{i\omega t_{d1}})\vert\varphi_2^\perp\rangle\rbrack\\
=&...
\end{split}
\end{equation}
where $a_i$, $b_i$ and $t_{di}$ (i=1,2,...) are real numbers, the values of which can be obtained from Eq.(\ref{eq:Bm}-\ref{eq:Aeff}) by setting $N=1$ and $m=i$, and $|\varphi_i\rangle=e^{-i\hat{A}\omega\tau_i}...e^{-i\hat{A}\omega\tau_1}|\varphi_{in}\rangle=e^{-i\hat{A}\omega\sum\tau_i}|\varphi_{in}\rangle$. As the cross-talk is small, i.e. $b_i\ll a_i$, so that the term $b_1b_2e^{i\omega t_{d1}}$ can be neglected. In consequence, the process $...e^{-i\hat{A}\omega\tau_2}M_2e^{-i\hat{A}\omega\tau_1}M_1$ is approximate equivalent to $...e^{-i\hat{A}\omega\tau_2}e^{-i\hat{A}\omega\tau_1}M_2M_1$. Similarly, we can equivalently put all $\{M_i\}$ together in front of all $\{e^{-i\hat{A}\omega \tau_i}\}$, as is shown in Fig.\ref{fig:amplitude_noise}. Finally, we can achieve the final output state in presence of amplitude-type noise:

\begin{equation}
|\varphi'_{out}\rangle=a|\varphi _{out}\rangle+be^{i\omega t_{d}}|\varphi^{\perp } _{out}\rangle,
\end{equation}
where $\varphi _{out}$ denotes the output state without noise, and $\varphi^{\perp } _{out}$ denotes the state being orthogonal to $\varphi _{out}$. The values of $a$, $b$ and $t_d$ can be obtained from Eq.(\ref{eq:Bm}-\ref{eq:Aeff}):
\begin{equation}
a=e^{ikn_{o}L}\prod_{m=1}^{N}B_{m}\quad\quad  b=A_{eff}\quad\quad  t_{d}=\frac{\theta_{eff}}{\omega}
\end{equation}

After the postselection, we can derive the pointer spectrum $P(\omega)=|\langle \phi_f|\varphi'_{out}\rangle|^2$. In all four schemes, $P_{0}(\omega)$ represents the spectrum without noise and $\left | f(\omega) \right |^{2}$ represents the spectrum of the broad-band source. 

For SWVA, the pointer spectrum after introducing the noise becomes
\begin{equation}\label{eq:p_swva}
\begin{aligned}
P(\omega)=&a^{2}P_{0}(\omega)+\left [b^{2}\cos^{2}(\omega\tau-\epsilon)
\right.
\\
\phantom{=\;\;}
&\left.-ab(\sin 2(\omega\tau-\epsilon))\cos(\omega t_{d})   \right ] \left | f(\omega) \right |^{2}
\end{aligned}
\end{equation}

For BWVA, the pointer spectrum after introducing the noise becomes
\begin{equation}\label{eq:p_bwva}
\begin{aligned}
P(\omega)=&a^{2}P_{0}(\omega)+\left [ b^{2}\cos^{2}(\omega\tau-\epsilon )
\right.
\\
\phantom{=\;\;}
&\left.+ab\sin2(\omega \tau -\epsilon )\sin(\omega t_{d}) \right ]\left | f(\omega) \right |^{2}
\end{aligned}
\end{equation}

For JWVA, the pointer spectrum after introducing the noise becomes
\begin{equation}\label{eq:p_jwva}
\begin{cases}
\begin{aligned}
P_{1}(\omega)=&P_{10}(\omega)+\left [ b^{2}\sin^{2}(\frac{\omega\tau+\epsilon }{2})
\right.
\\
\phantom{=\;\;}
&\left.-ab\sin(\omega\tau+\phi)\sin(\omega t_{d}) \right ]\left | f(\omega) \right |^{2}
\\
P_{2}(\omega)=&P_{20}(\omega)+\left [ b^{2}\sin^{2}(\frac{\omega\tau+\phi }{2})
\right.
\\
\phantom{=\;\;}
&\left.+ab\sin(\omega\tau+\phi)\sin(\omega t_{d}) \right ]\left | f(\omega) \right |^{2}
\end{aligned}
\end{cases}
\end{equation}

For DWVA, the pointer spectrum after introducing the noise becomes
\begin{equation}\label{eq:p_dwva}
P_{\pm }(\omega)=P_{0\pm }(\omega)+\left | f(\omega) \right |^{2}
\left \{ P_{Da}+P_{Db}\mp P_{Dc}\right \},
\end{equation}

where
\begin{equation}
\begin{cases}
P_{Da}=\frac{1}{2}\left \{ 1\mp \sin2[(1-\frac{\omega}{\omega_{0}})\epsilon +\tau\omega ]\right \},\\
P_{Db}=\left \{ 
\frac{\sqrt{2}}{2}\sin(\omega t_{d}-\frac{\pi}{4}))(\pm1+ \sin2[(1-\frac{\omega}{\omega_{0}})\epsilon +\tau\omega]) 
\right \},\\
P_{Dc}=\left \{ 
\frac{\sqrt{2}}{2}\sin(\omega t_{d}+\frac{\pi}{4}))(\cos2[(1-\frac{\omega}{\omega_{0}})\epsilon +\tau\omega]) 
\right \}.
\end{cases}
\end{equation}

From Eq.(\ref{eq:p_swva}-\ref{eq:p_dwva}), we can see that after inducing the amplitude-type noise sine-function-like terms are involved, therefore the average frequency shift $\delta_{\omega}$ obtained by Eq.(\ref{eq:delta_omega}) are different from the ones that without noises. Consequently, the sensitivities of the four WVA approaches are reduced comparing to that obtained by Eq.(\ref{eq:sensitivity_ideal}).

To study the sensitivity reduction, we perform numerical simulations based on Eq.(\ref{eq:p_swva}-\ref{eq:p_dwva}) and Eq.(\ref{eq:delta_omega}). We assume that the initial spectrum of the light source to be Gaussian, with center optical wavelength (angular frequency) of $1550nm$ ($1.216\times 10^{15}Hz$) and bandwidth of $10nm$ ($7.846\times 10^{12}Hz$). We first calculate the relation between $\delta_{\omega}$ and $\tau$, and the corresponding sensitivity can be obtained as the maximum value of $d\delta_{\omega}/d\tau$, as was defined before. 

\begin{figure}[ht!]
\centering
    \subfigure[]{\includegraphics[width=0.32\textwidth]{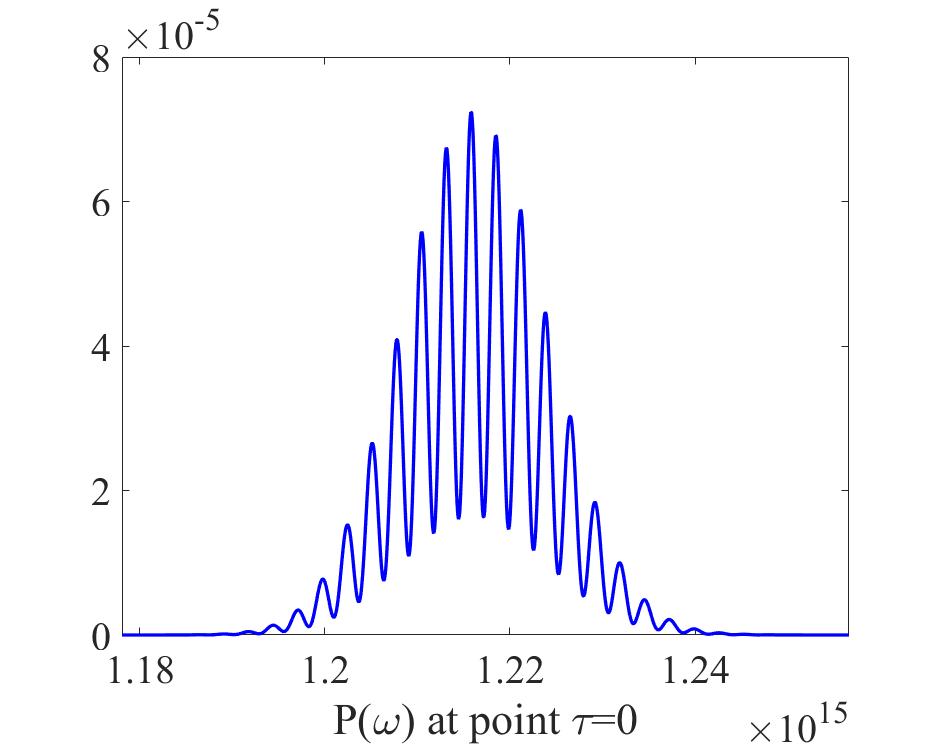}} 
    \subfigure[]{\includegraphics[width=0.32\textwidth]{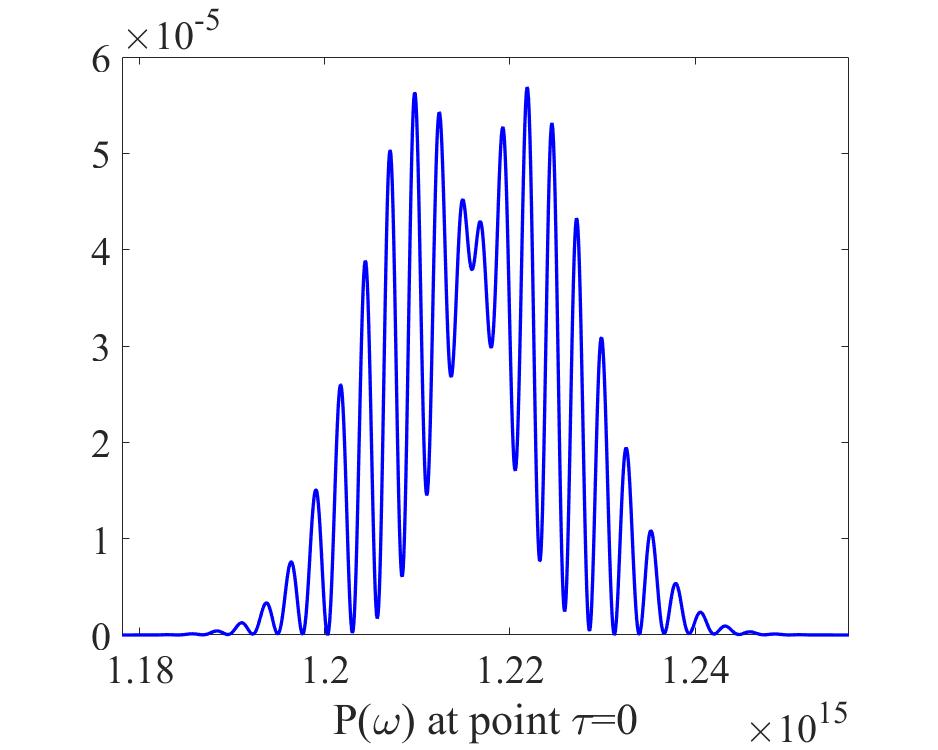}} \\
    \subfigure[]{\includegraphics[width=0.32\textwidth]{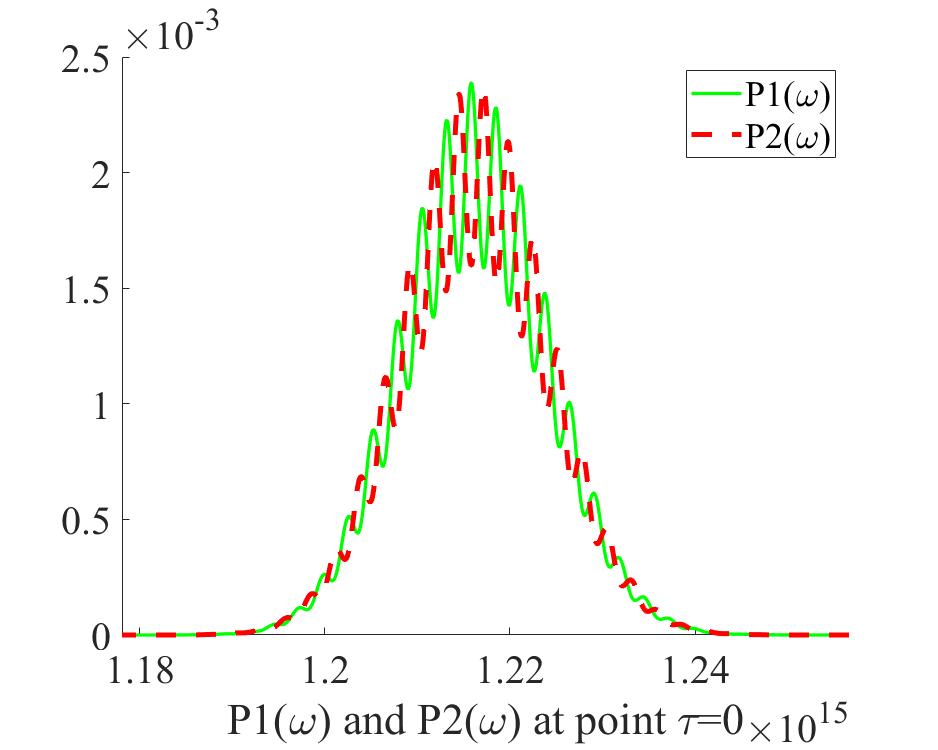}} 
    \subfigure[]{\includegraphics[width=0.32\textwidth]{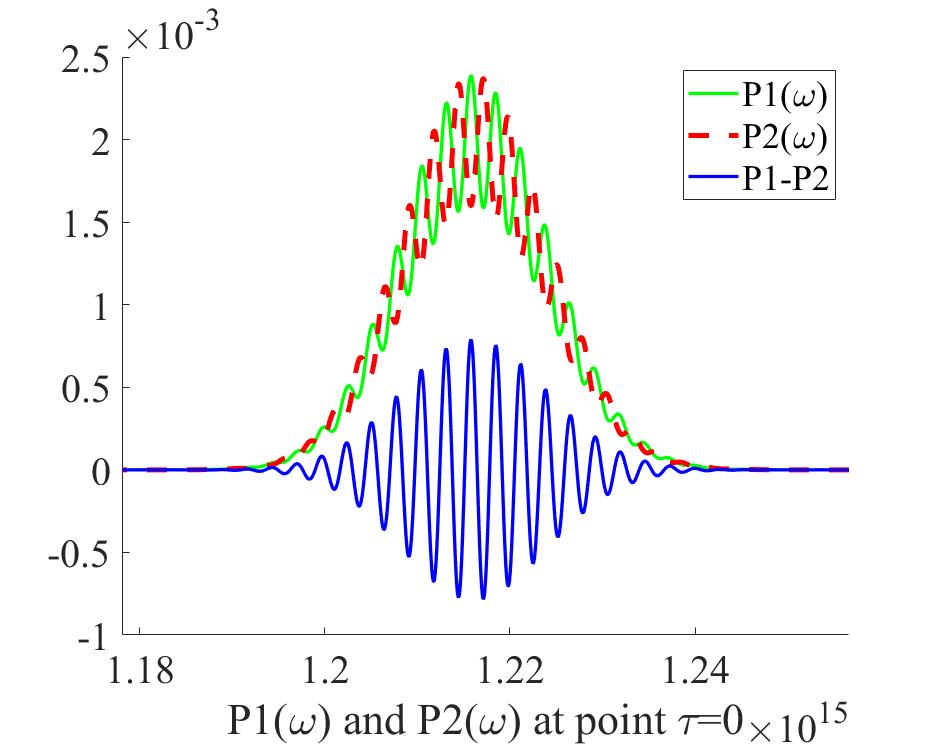}} \\
\caption{Numerical simulations on the spectrum of WVA approaches in presence of amplitude-type noise ($|b|/|a|=0.1$): (a)SWVA, (b)BWVA, (c)JWVA, (d)DWVA.}
\label{simuspectrums}
\end{figure}

\begin{figure}[ht!]
\centering
    \subfigure[]{\includegraphics[width=0.32\textwidth]{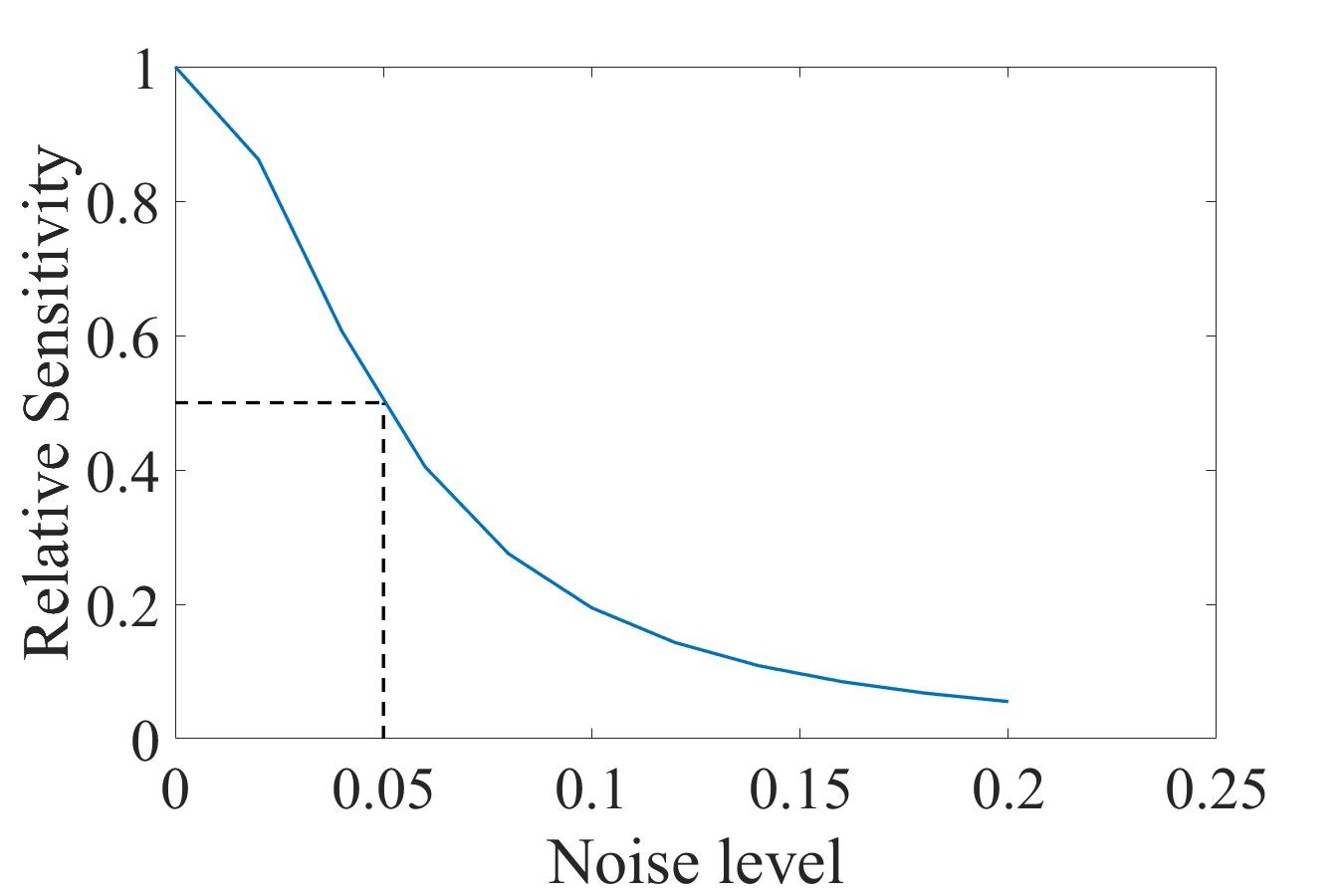}} 
    \subfigure[]{\includegraphics[width=0.32\textwidth]{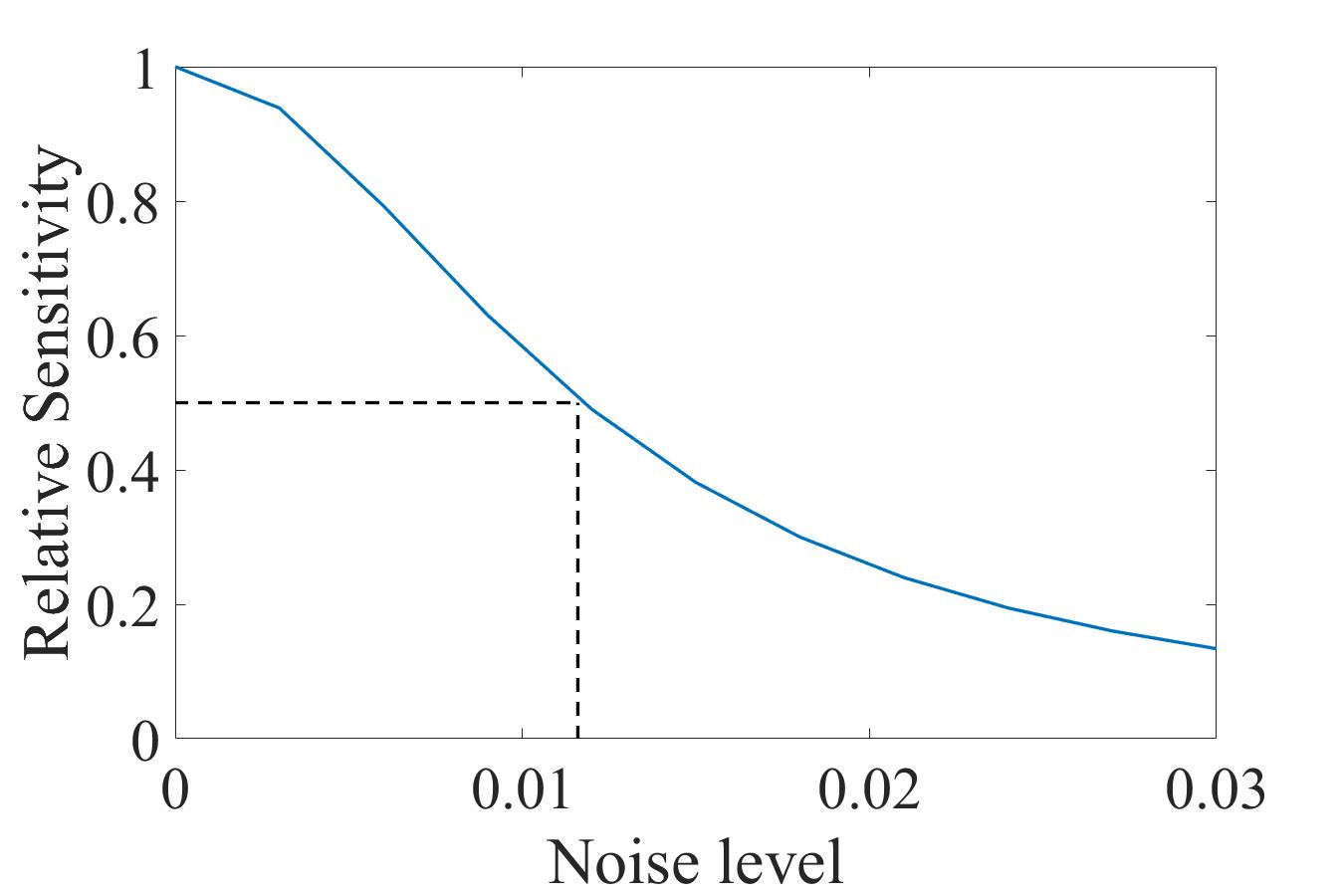}} \\
    \subfigure[]{\includegraphics[width=0.32\textwidth]{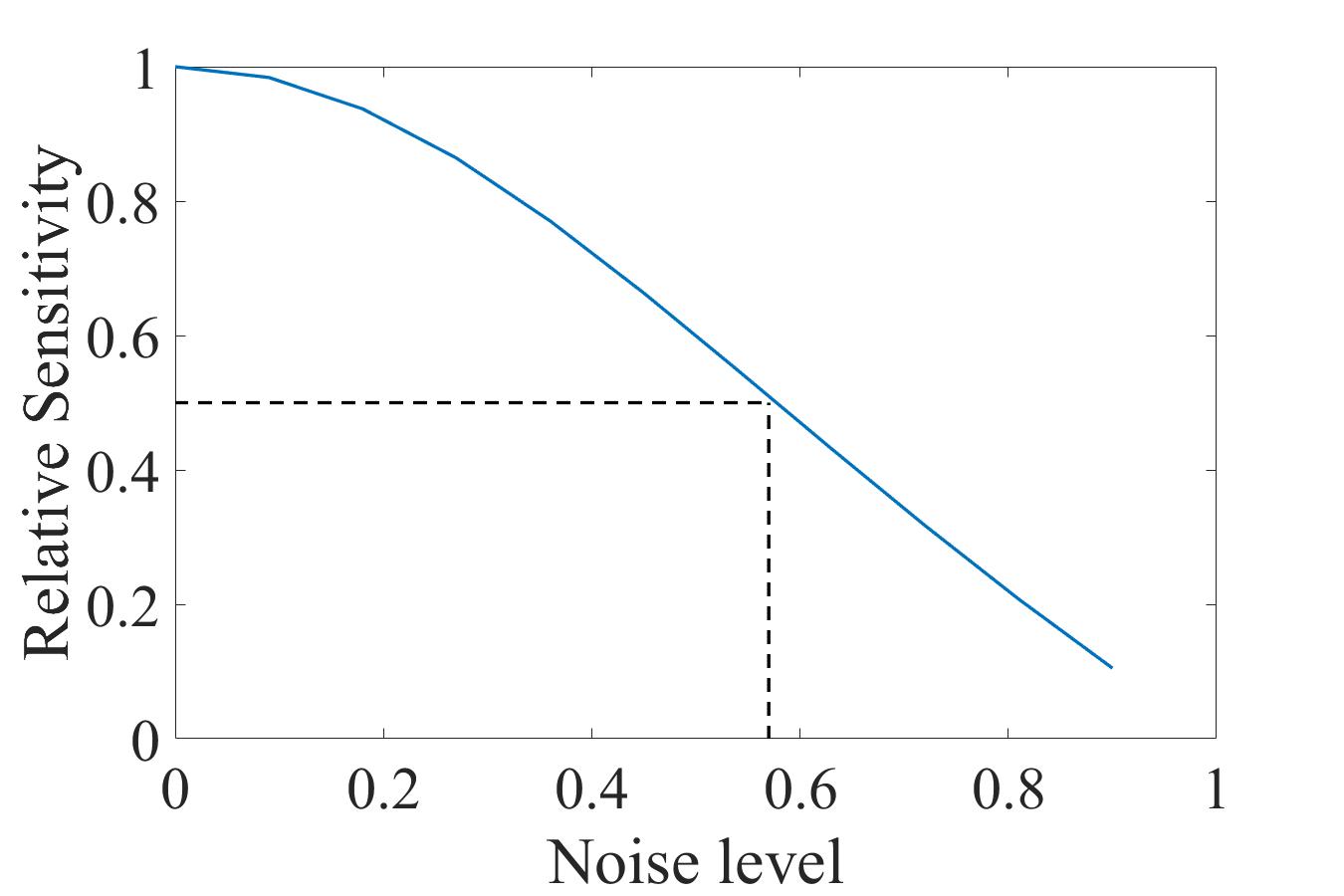}} 
    \subfigure[]{\includegraphics[width=0.32\textwidth]{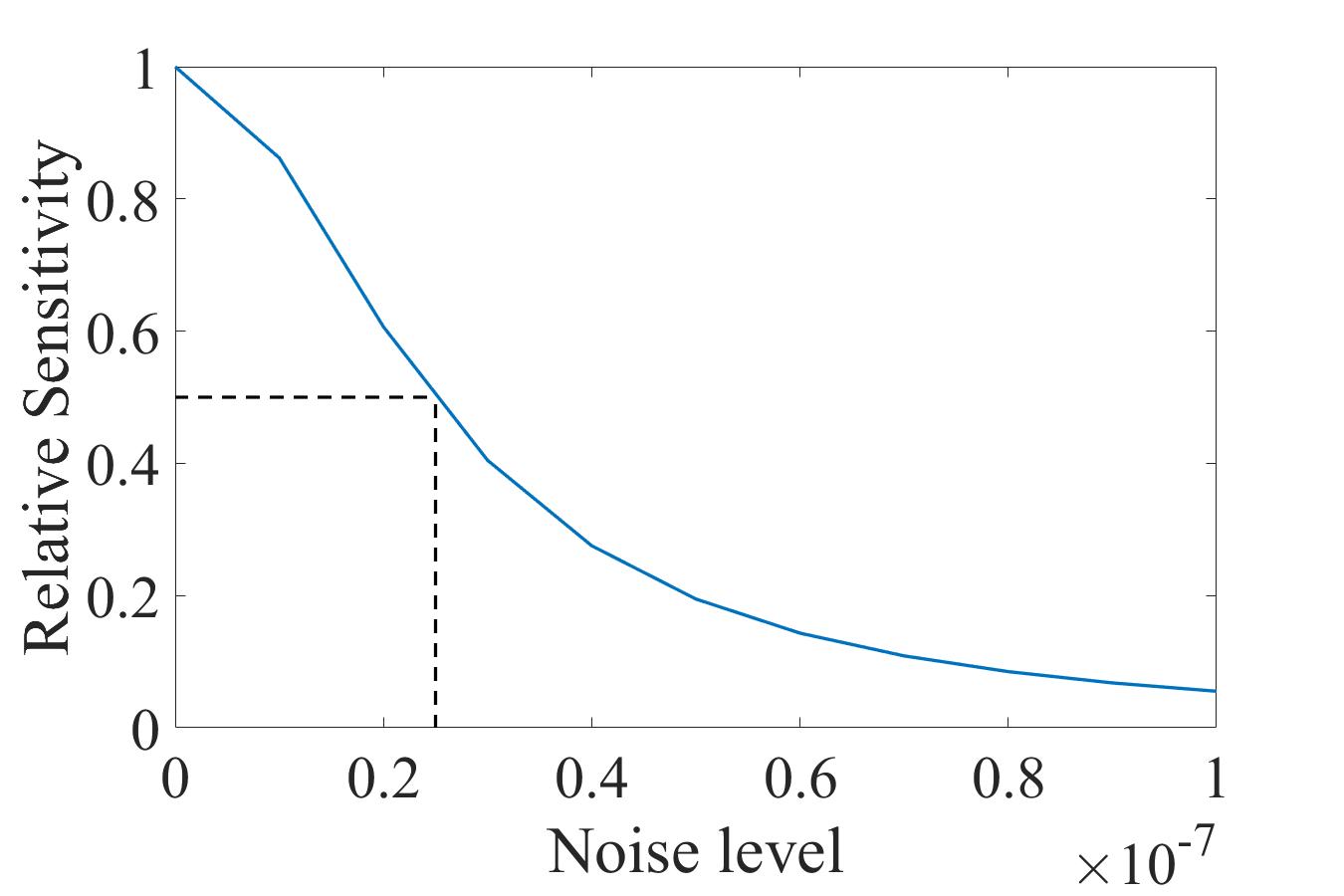}} \\
\caption{Numerical simulations on the relative sensitivities (with respect to the sensitivities without noise) of WVA approaches: (a)SWVA, (b)BWVA, (c)JWVA, (d)DWVA.}
\label{simucurves}
\end{figure}

The output spectra of different WVA approaches in presence of amplitude-type noise are given in Fig.\ref{simuspectrums}, where $t_d$ is assumed to be $7.0 \times 10^{-4}s$. Comparing to the spectrum without noise shown in Fig.\ref{WVANoNoise}, we can see that the shapes of spectrum are obviously distorted. While the broadband optical spectrum is seriously interfered by the amplitude-type noise here, the effect cannot be mitigated by narrowing the line width of the source. Moreover, the relation between the sensitivity and the relative noise ratio for each approach  is shown in Fig.\ref{simucurves}, where the relative noise ratio is defined as $|b|/|a|$. Apparently, amplitude-type noise will seriously reduce the sensitivity. 


As is summarized in Table \ref{tab1}, the relative noise intensities corresponding to a $3 dB$ sensitivity loss are $0.05$ and $0.55$ for SWVA and JWVA respectively, and rapidly drop down to $1.1\times10^{-2}$ and $2.5\times10^{-8}$ for BWVA and DWVA. Therefore although BWVA and DWVA can provide much higher sensitivities comparing to SWVA and JWVA\cite{Huang2019}, their robustness against amplitude noise is another story. In any case, the amplitude-type noise will significantly compromise the performance of WVA, so that it is important to suppress this type of noise in the WVA experiments using optical fiber. 

\begin{table}[ht!]
\newcommand{\tabincell}[2]{\begin{tabular}{@{}#1@{}}#2\end{tabular}}
\centering
\begin{scriptsize}
\caption{\label{tab1}\textbf{Relative amplitude-type noise ratio with $ 3dB $ sensitivity loss}}
\scalebox{1.0}
{
\begin{tabular}{ccccc} \hline
  & SWVA & BWVA & JWVA & DWVA \\ \hline
\tabincell{c}{Relative amplitude-type\\ noise ratio with $ 3dB $\\ sensitivity loss} & $5.0\times10^{-2}$ & $1.1\times10^{-2}$
 & $5.5\times10^{-1}$ & $2.5\times10^{-8}$  \\\hline
\end{tabular}
}
\end{scriptsize}
\end{table}

\section{Experimental results}
\subsection{Experiment setup and results}

\begin{figure}[ht!]
    \centering
    \includegraphics[width=8.5cm]{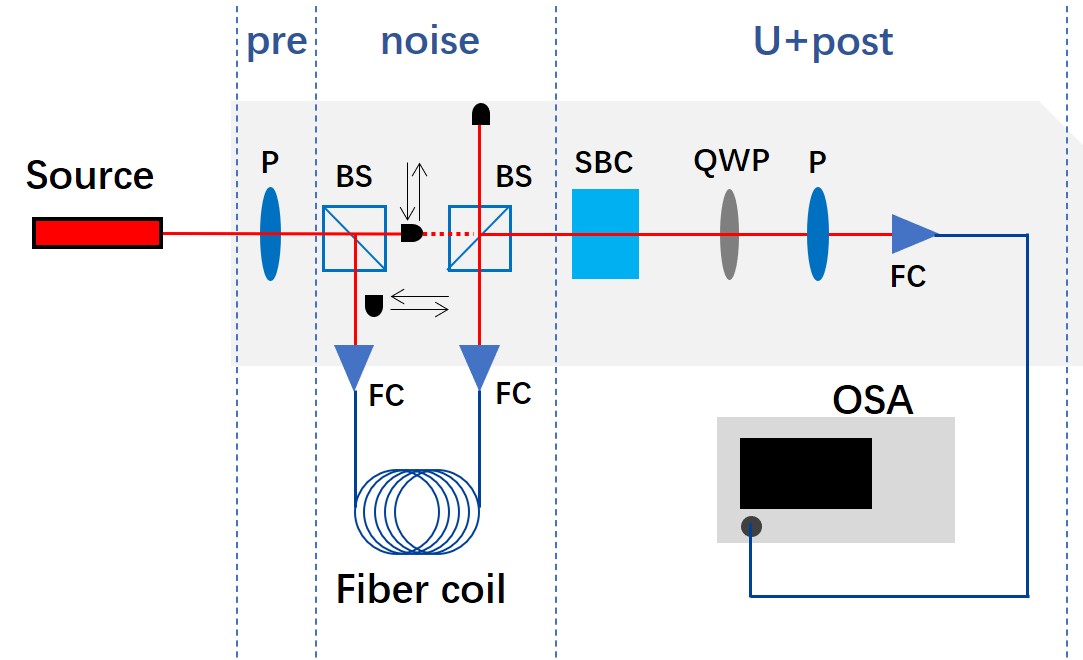}
    \caption{Experimental setup for demonstrating the effect of amplitude-type noise in BWVA scheme. Source: Superluminescent Diode, P:polarizer which OAs of first and later one aligned with angle of $\frac{\pi}{4}$ an $-\frac{\pi}{4}+\epsilon$ respectively, BS:beam splitter, FC:fiber collimator,  SBC:Soleil-Babinet Compensator on direction of $\frac{\pi}{4}$, QWP: quarter wave plate which OA set at the angle $-\frac{\pi}{4}$, OSA:optical spectrum analyzer.}
    \label{BWVA Exp Setup}
\end{figure}

In this section, we analyze the effect of the amplitude-type noise in optic-fiber-based WVA by experiment, where the BWVA scheme (with experimental setup shown in Fig.\ref{BWVA Exp Setup}) is taken as example. In the experiment, a light beam with central wavelength of $\lambda_0=1550 nm$ and bandwidth of $10nm$ is generated by the superluminescent diode (SLD), and the initial polarization state on $45^{\circ}$ direction is modulated by a linear polarizer. The light is then decided to pass through free-space or a 1-meter long coiled polarization maintaining fiber (PMF), by blocking one of the two paths behind a beam splitter (BS). The processes of interaction and post-selection consist of a Soleil-Babinet compensator (SBC), a quarter wave plate (QWP), and another linear polarizer. Here, the optical axis of the SBC is tilted by angle of $\pi/4$. The SBC is applied to modulate a longitudinal phase change of $\omega(\epsilon\lambda_0/2\pi c + \tau)$ between the polarizations H and V, where the time-delay $\tau$ is the parameter of interest in the experiment.\cite{Huang2019}. After that, the light passes the postselection part consisting of a QWP and the second linear polarizer, in which the axes of the QWP and the polarizer are set at angles of $-\pi/4$ and $-\pi/4 + \epsilon$. Finally, the spectrum of output light is measured by the optical spectrum analyzer (OSA).

\begin{figure}[ht!]
    \centering
    \includegraphics[width=8cm]{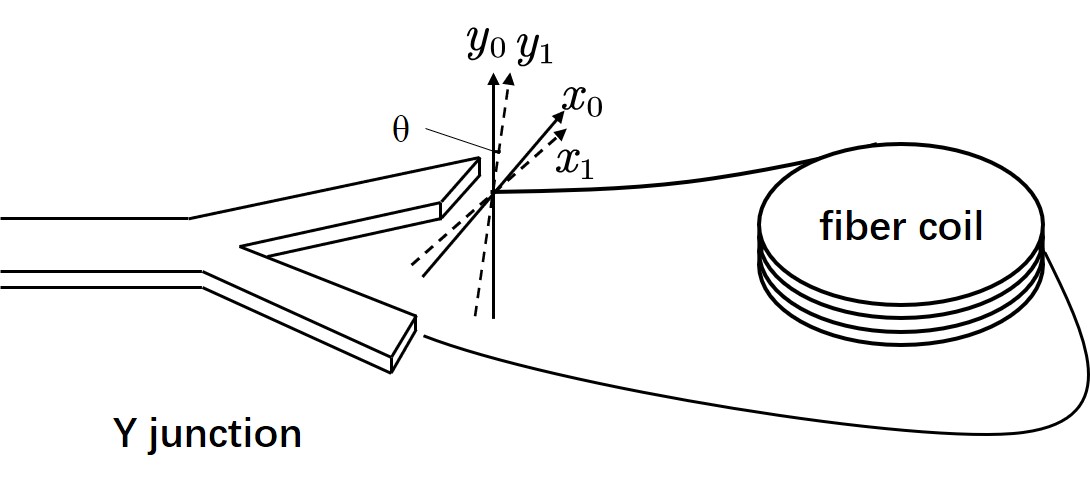}
    \caption{Y-junction and fiber coil in the fiber-optic gyroscope.}
    \label{Y-junction}
\end{figure}

In this experiment, the amplitude-type noise is induced by the misalignment between the directions of the incident light polarization and the fast axis of the optical fiber, which imitates the main reason of originating amplitude-type noise in optical fiber sensors\cite{lefevre2014fiber}. 
For example, in a fiber-optic gyroscope, the optical axes of the Y-junction and optical fiber are inevitably misaligned by an angle of $\theta$\cite{lefevre2014fiber} (see Fig.\ref{Y-junction}). The corresponding amplitudes of the noise and signal (which are denoted as $A_{amp}$ and $A_{signal}$ respectively) are related by $A_{amp}/A_{signal}=\tan(\theta)$.

\begin{figure}[ht!]
    \centering
    \subfigure[]{\includegraphics[width=0.42\textwidth]{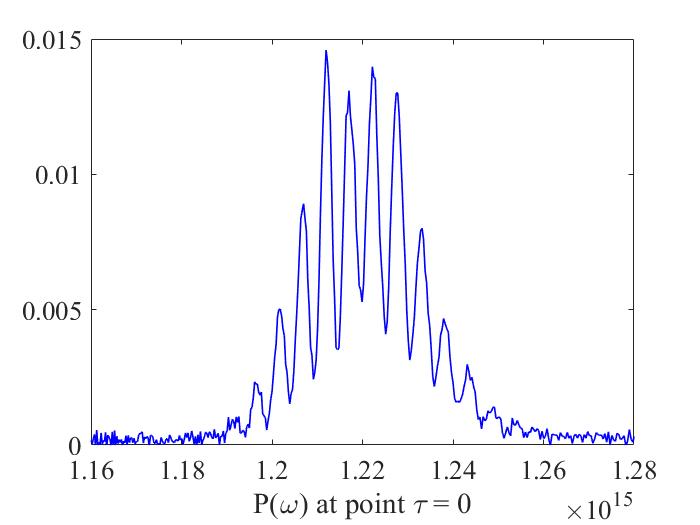}} 
    \subfigure[]{\includegraphics[width=0.42\textwidth]{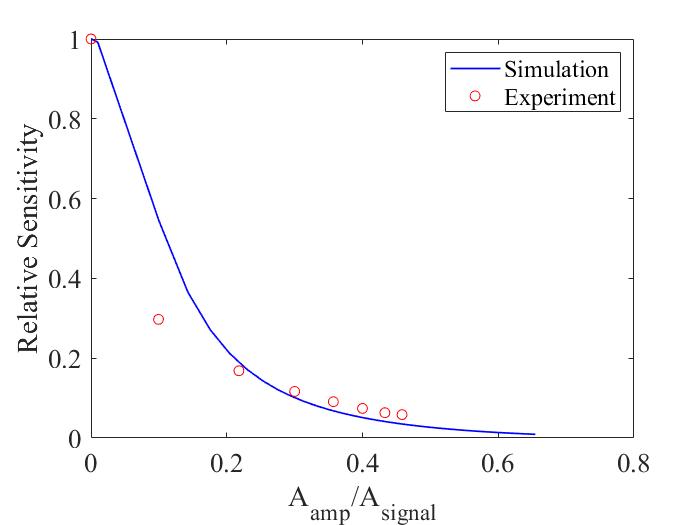}} \\
    \caption{Experimental and simulation results: (a)The light spectrum measured from the optical fiber path. (b)The relative sensitivities based on center wavelength (with respect to the sensitivities measured from the free-space path) of the optical fiber path, simulation (in blue line) and experimental (in red points).}
    \label{BWVA Experiment}
\end{figure}

Main results of the experiment are shown in Fig.\ref{BWVA Experiment}. From Fig.\ref{BWVA Experiment}(a), we can find that the light spectrum measured from the optical fiber path is different from the numerical simulation given in Fig.\ref{simuspectrums}(b). The reason is that in the experiment, in addition to the amplitude-type noise, there is unavoidable intensity noises introduced by a number of factors, such as equipment defects, the low extinction ratio of the wave plates, decoherence of light, multiple reflection/scattering in the experiment setup and so on\cite{Li2019}. Using the method we had derived in Ref\cite{Li2019}, the proportion of intensity noise is estimated to be $20\%$. According to this consideration, modified simulation of relative sensitivity is present along with the experimental results in Fig.\ref{BWVA Experiment}(b). 
When $A_{amp}/A_{signal}$ increases, the relative sensitivity decreases accordingly, and the experimental results are in good agreement with the theoretical predictions. 

The angle between the fast axis of the fiber coil and the polarization direction of the incident light should be alighed as small as possible to retain higher sensitivity. In practice, this angle can be made as small as $1^{\circ}$, corresponding to a relative sensitivity of 0.9553 when intensity noise is not taken into account. For a requirement of the sensitivity dropping no more than $3dB$, the value of $A_{amp}/A_{signal}$ is required to be smaller than 0.08, which means the misalignment error should be made smaller than $0.08 rad$.

\subsection{Discussion}

It was suggested that the sensitivity of WVA is not limited by the spectral resolution of the OSA, therefore the full-scan spectrum analysis can be take place by a split detection on the spectrum\cite{2020JPhB...53u5501H}. As the later one is much simpler for implementation, it would be interesting to testify whether the conclusion derived in Ref.\cite{2020JPhB...53u5501H} still available in presence of amplitude noise. To this goal, we divide the output light spectrum into two parts: the part of $\omega < \omega_0$ and the other part of $\omega > \omega_0$, where $\omega_0$ is the average optical frequency of the initial light spectrum. The intensities of these parts are then accumulated, normalized, and finally subtracted, it ideally yields:
\begin{equation}
\begin{split}
    \Delta I&=\frac{I_2-I_1}{I_1+I_2}=\frac{\int_{\omega_0}^\infty P(\omega)\operatorname d\omega-\int_0^{\omega_0}P(\omega)\operatorname d\omega}{\int_0^{\omega_0}P(\omega)\operatorname d\omega+\int_{\omega_0}^\infty P(\omega)\operatorname d\omega}\\
    &=\frac{-i\sin2(\omega_0\tau-\varepsilon)\mathrm{erf}(i\sigma\tau)}{\mathrm e^{\sigma^2\tau^2}-\cos2(\omega_0\tau-\varepsilon)}.\;
\end{split}
\end{equation}
When the amplitude noise is taken into consideration, the formula is modified to be:
\begin{equation}
\Delta I'=\frac{I_2'-I_1'}{I_1'+I_2'}
\end{equation}
where
\begin{equation}
\begin{aligned}
{I_2'-I_1'}=&-i(a^2-b^2)\sin2(\omega_0\tau-\varepsilon)\mathrm{erf}(i\sigma\tau)-iabe^{-\sigma^2({\textstyle\frac14}t_d^2+\tau t_d)}\cdot\\
&\cdot\sin2\lbrack\omega_0(\tau+{\textstyle\frac12}t_d)-\varepsilon\rbrack\cdot\mathrm{erf}\lbrack i\sigma(\tau+{\textstyle\frac12}t_d)\rbrack+\\
&iabe^{-\sigma^2(\frac14t_d^2-\tau t_d)}\sin2\lbrack\omega_0(\tau-\frac12t_d)-\varepsilon\rbrack\mathrm{erf}\lbrack i\sigma(\tau-\frac12t_d)\rbrack
\end{aligned}
\end{equation}
and
\begin{equation}
\begin{aligned}
{I_1'+I_2'}=&(a^2+b^2)e^{\sigma^2\tau^2}-(a^2-b^2)\cos2(\omega_0\tau-\varepsilon)\\
&-abe^{-\sigma^2(\frac14t_d^2+\tau t_d)}\cos2\lbrack\omega_0(\tau+\frac12t_d)-\varepsilon\rbrack\\
&+abe^{-\sigma^2(\frac14t_d^2-\tau t_d)}\cos2\lbrack\omega_0(\tau-\frac12t_d)-\varepsilon\rbrack
\end{aligned}
\end{equation}

\begin{figure}[ht!]
\centering
    \subfigure[]{\includegraphics[width=0.42\textwidth]{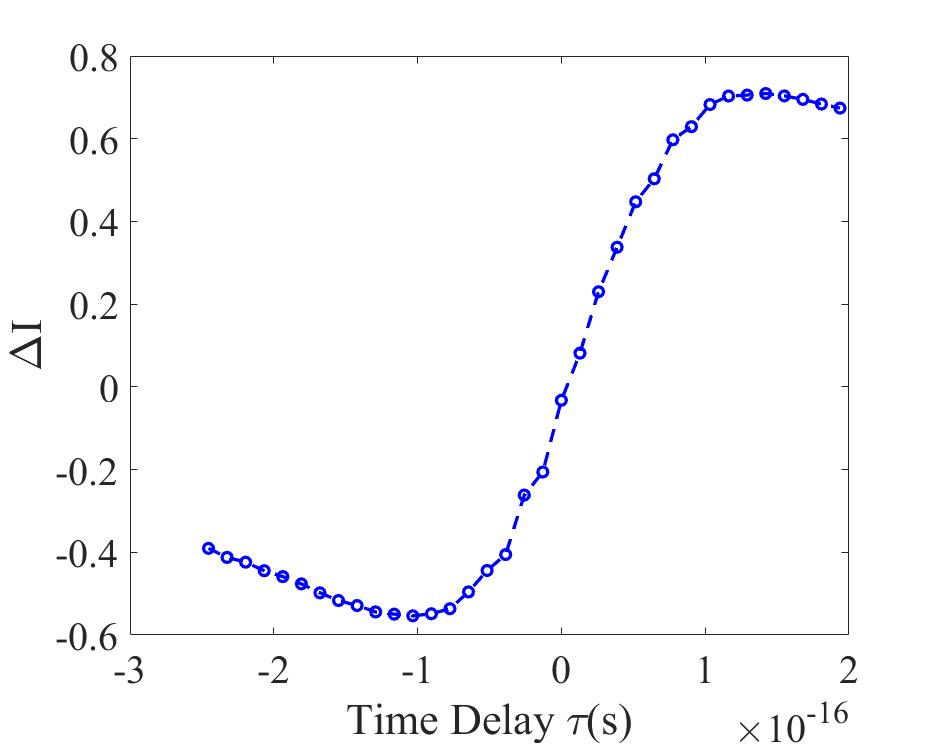}} 
    \subfigure[]{\includegraphics[width=0.42\textwidth]{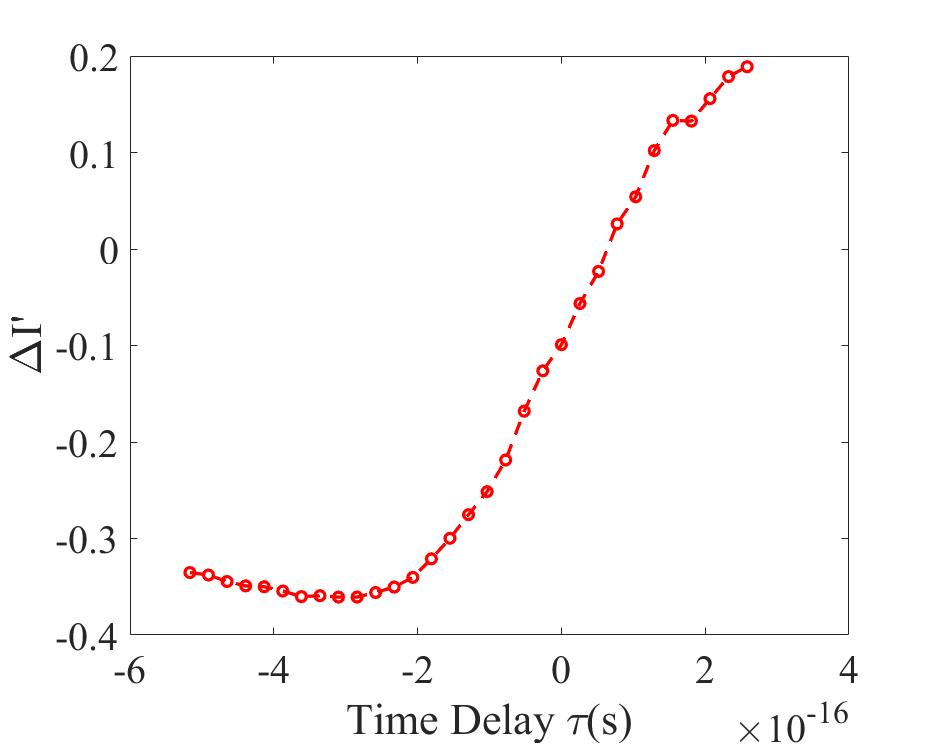}} \\
\caption{ $\Delta I$ calculated based on the experimental data obtained from (a) the free-space path (without amplitude-type noise) and (b) the optical fiber path (with amplitude-type noise).}
\label{DividedSpectrumResult}
\end{figure}

\begin{figure}[ht!]
\centering\includegraphics[width=7cm]{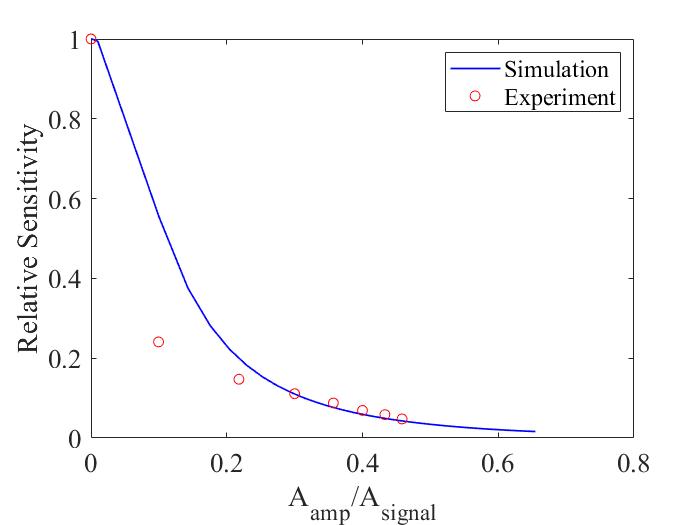}
\caption{The relative sensitivity of optical-fiber path with respect to the free-space path, utilizing the spectral split detection scheme. The blue line is the numerical simulation and red circles are experimental results.}
\label{Divided spectrum}
\end{figure}

The results of the simulation and experiment are shown in Fig.\ref{DividedSpectrumResult}(a) and Fig.\ref{Divided spectrum}. Fig.\ref{DividedSpectrumResult}(a) and (b) respectively show the relation between $\tau$ and $\Delta I$ without and with the amplitude-type noise. The sensitivity decreases when amplitude-type noise is in present, which is similar to the case of using full-scan spectrum analysis.
More precisely, by comparing the changing relative sensitivity in split detection scheme (see Fig.\ref{Divided spectrum}) the one in full-scan scheme (see Fig.\ref{BWVA Experiment}(b)), we can find that their dependency on $A_{amp}/A_{signal}$ is nearly the same, which means the split detection scheme can tolerate the amplitude-type noise as well as the full-scan scheme. 
In consequence, spectral split detection can be applied in the fiber-optic-based WVA system to replace the OSA, making the detection process more simplified and efficient.

\section{Conclusion}
In conclusion, we theoretically and experimentally studied the effect of amplitude-type noise induced by optical fiber in the WVA schemes using spectral analysis. By exemplifying all of the four approaches, we find that the BWVA approach is robust to this optical-fiber induced imperfection. In order to overcome the effect of amplitude-type noise, the angular alignment error at the interface should be made as small as possible. According to our experiment, the misalignment should be smaller than $0.08 rad$ once the sensitivity drop down is required to be no more than $3dB$. 
Moreover, we testify the spectral split detection method in presence of amplitude-type noise, and find that it can tolerate this kind of noise as well, which means this simpler detection can take place of the complex OSA to make the optical-fiber based system simpler and more efficient.
The current results indicate that implementing WVA based on optical fiber is feasible, and it provides a new way for designing optical sensors that acquiring the technical advantages from WVA (high sensitivity) and optical fiber (high stability) simultaneously. Furthermore, the impacts of polarization mode dispersion(PMD) and polarization dependent loss(PDL) in optical fiber are also important and worthy for further investigations\cite{Zhao:22}.

\begin{backmatter}
\bmsection{Funding} National Natural Science Foundation of China (62071298, 61901258), Civil Aerospace Advance Research Project (D020403), and the fund of State Key Laboratory of Advanced Optical Communication Systems and Networks.

\bmsection{Disclosures} The authors declare that there are no conflicts of interest related to this article.

\bmsection{Data availability} Data underlying the results presented in this paper are not publicly available at this time but may be obtained from the authors upon reasonable request.

\end{backmatter}




\bibliography{sample}



\ifthenelse{\equal{\journalref}{aop}}{%
\section*{Author Biographies}
\begingroup
\setlength\intextsep{0pt}
\begin{minipage}[t][6.3cm][t]{1.0\textwidth} 
  \begin{wrapfigure}{L}{0.25\textwidth}
    \includegraphics[width=0.25\textwidth]{john_smith.eps}
  \end{wrapfigure}
  \noindent
  {\bfseries John Smith} received his BSc (Mathematics) in 2000 from The University of Maryland. His research interests include lasers and optics.
\end{minipage}
\begin{minipage}{1.0\textwidth}
  \begin{wrapfigure}{L}{0.25\textwidth}
    \includegraphics[width=0.25\textwidth]{alice_smith.eps}
  \end{wrapfigure}
  \noindent
  {\bfseries Alice Smith} also received her BSc (Mathematics) in 2000 from The University of Maryland. Her research interests also include lasers and optics.
\end{minipage}
\endgroup
}{}

\end{document}